\documentclass[10pt,journal,twocolumn,twoside]{IEEEtran}
\usepackage{setspace}
\usepackage{cite}
\usepackage{color}
\usepackage{amsmath}
\usepackage{amsfonts}
 \usepackage{amssymb}
\usepackage{graphicx}
\usepackage{algorithm}
\usepackage{algorithmic}
\usepackage{cleveref}
\usepackage{url}
 \usepackage[noblocks]{authblk}
 \usepackage{makecell}
 \usepackage{verbatim}
\usepackage{booktabs}

\begin{document}

\title{Enhanced Fluid Index Modulation for Integrated Data and Energy Transfer}

\author{Long Zhang,~\IEEEmembership{Graduate Student Member,~IEEE}, Yizhe Zhao,~\IEEEmembership{Member,~IEEE}, Halvin Yang, \IEEEmembership{Member, IEEE}, Qiang Liu,~\IEEEmembership{Senior Member,~IEEE}, Kai-Kit Wong,~\IEEEmembership{Fellow,~IEEE}

\thanks{ Long Zhang and Yizhe Zhao are with the School of Information and Communication Engineering, University of Electronic Science and Technology of China, Chengdu 611731, China (e-mail: l.zhang@std.uestc.edu.cn; yzzhao@uestc.edu.cn). }

\thanks{Halvin Yang is  with the Department of Electrical and Electronic Engineering, Imperial College London, SW7 2AZ London, U.K. (e-mail: halvin.yang@imperial.ac.uk).}

\thanks{Qiang Liu is with the Yangtze Delta Region Institute (Quzhou), University of Electronic Science and Technology of China, Quzhou 324000, Zhejiang, China (e-mail: liuqiang@uestc.edu.cn).}

\thanks{Kai-Kit Wong is affiliated with the Department of Electronic and Electrical Engineering, University College London, Torrington Place, WC1E 7JE, United Kingdom and he is also affiliated with the Department of Electronic Engineering, Kyung Hee University, Yongin-si, Gyeonggi-do 17104, Korea (e-mail: kai-kit.wong@ucl.ac.uk).}
}

\maketitle

\vspace{-2cm}

\begin{abstract}

	Integrated data and energy transfer (IDET) is a promising technique for supporting sustainable low-power wireless networks. To improve both communication reliability and energy transfer efficiency, this paper investigates a fluid index modulation (FIM) assisted IDET system, where the base station employs a two-dimensional fluid antenna system (FAS) and the receiver adopts a power-splitting architecture. In FIM, the information bits are delivered not only from the modulation symbols, but also the index of antenna position. Under finite-alphabet signaling, the average harvested power, bit error rate (BER), and achievable data rate are derived in closed form. A joint optimization problem is formulated to maximize the average harvested power subject to BER and achievable rate constraints by jointly optimizing the port selection, precoding vector, and power splitting ratio. An alternating optimization framework is developed, where the precoding vector and port selection are obtained via a Riemannian augmented Lagrangian method (RALM) and block coordinate descent (BCD) algorithm, respectively. Simulation results demonstrate that the proposed scheme achieves a superior rate-energy trade-off over benchmark schemes, while the proposed algorithm attains near-optimal performance with significantly lower complexity than exhaustive search.
\end{abstract}

\begin{IEEEkeywords}
Integrated data and energy transfer (IDET), fluid index modulation (FIM), port selection, precoding design.
\end{IEEEkeywords}

%
\IEEEpeerreviewmaketitle

\section{Introduction}
\subsection{Background}
With the rapid evolution of sixth-generation (6G) networks and the ubiquitous deployment of the Internet of Things (IoT), massive wireless devices are expected to be connected in future communication systems \cite{10858129}. This trend places increasing pressure on both resource utilization and device energy supply. Notably, most IoT devices are characterized by compact size, limited hardware capabilities, and strict energy budgets, making their continuous operation heavily reliant on finite battery capacities. Consequently, repeated battery charging or replacement not only raises maintenance cost, but also limits the long-term and large-scale deployment of such networks. In response to the above challenges, integrated data and energy transfer (IDET) has been proposed as a potential solution, as radio frequency (RF) signals can simultaneously support wireless data transfer (WDT) and wireless energy transfer (WET) \cite{9675011}. In practical IDET systems, the received RF signal is usually divided into two parts at the receiver, either in the time domain or in the power domain. One part is used for energy harvesting, while the other is used for information decoding. Through such a signal splitting mechanism, the receiver can flexibly balance WDT and WET according to different system requirements. However, the performance of IDET is often limited by the severe attenuation and fading of RF signals over wireless channels, which results in low WET efficiency. In this regard, multiple-input-multiple-output (MIMO) technology has been introduced to enhance both information transmission and energy transfer by exploiting spatial degrees of freedom. Nevertheless, the antenna positions in conventional MIMO systems are fixed, which limits their ability to fully exploit the spatial variations of wireless channels within the available antenna aperture. 

Recently, fluid antenna system (FAS) has been proposed to enhance wireless communication performance via software-controlled reconfiguration of antenna positions \cite{11297439,10753482,11433651,9264694,11397315,11308117,11184548}. By enabling a single antenna element to switch among multiple ports within a compact region, FAS introduces additional spatial degrees of freedom without requiring multiple RF chains. In contrast to conventional Multiple-Input Multiple-Output (MIMO) systems, which rely on multiple fixed antennas and increased hardware complexity, FAS exploits the spatial variability of wireless channels to achieve diversity gains through port switching. As a result, it can effectively enhance the received signal strength and improve link reliability under stringent hardware and space constraints. Such capability is particularly beneficial for IDET systems, where both information decoding and energy harvesting critically depend on the received signal power. By adaptively selecting the port with favorable channel conditions, FAS can simultaneously improve the achievable data rate and increase the harvested energy. Therefore, the spatial reconfigurability of FAS naturally enables an efficient trade-off between wireless data transmission and wireless energy transfer, making it a promising solution for energy-constrained wireless networks.

Nevertheless, WET inevitably occupies extra resources, which in turn impairs the performance of WDT. Therefore,  index modulation (IM) has emerged as an attractive transmission technique for improving spectral efficiency and resource utilization \cite{8765384,9184265,10491109}. Unlike conventional amplitude-phase modulation, IM conveys additional information bits through the indices of communication resources, such as antennas, subcarriers, time slots, and channel states. In conventional fixed-antenna systems, however, implementing IM usually requires a relatively large number of antennas or other indexed resources, which inevitably increases hardware cost and system complexity. By contrast, fluid antenna systems naturally provide a large number of candidate ports within a compact aperture, making them particularly well suited for IM. In this way, the abundant ports of FAS can be exploited as index carriers without requiring a proportional increase in RF chains or hardware modules. Motivated by the complementary advantages of FAS and IM, integrating fluid antennas with index modulation offers a flexible and cost-effective design framework for IDET systems.

\subsection{Related Works}

IDET has attracted significant attention as a promising solution for sustainable low-power wireless networks. Existing studies on IDET mainly focus on transceiver design to enhance the rate-energy (R-E) trade-off between WDT and WET.  In \cite{10634829}, a nonlinear rectifier model was proposed for IDET systems, and the corresponding resource allocation was optimized for different receiver architectures. Zhang \textit{et al.} \cite{10500404} considered a practical IDET  scenario, where the data receivers (DRs) and energy receivers (ERs) were located in the far- and near-field regions of the extremely large-scale array (XL-array) base station (BS), respectively. The harvested weighted sum-power of DRs is maximized by jointly designing the beamforming and power allocation. In \cite{11260505}, the outage probability and asymptotic outage probability are analyzed in XL-array assisted multi-input single-output (MISO) IDET system. In \cite{10715704},  a hybrid time-switching/power-splitting protocol is considered in massive MIMO intelligent reflective surfaces (IRSs)-assisted IDET system, where the IRS phase-shifts and transmit power are jointly optimized to maximize the worst harvested energy and worst achievable rate. The work in \cite{10115291} considered a joint beamforming design at the BS and IRS, where the transmit power was minimized while guaranteeing the required DR and ER.
However, most existing IDET schemes are still built upon conventional fixed-position antenna (FPA) arrays. As a result, their performance remains fundamentally constrained by the limited spatial degrees of freedom offered by fixed antenna geometries, especially in rich-scattering and highly dynamic wireless environments.

Recently, some researchers have investigated FAS-assisted IDET systems to improve both WDT and WET performance \cite{10916605,10980171,11184149,10930949,10506795,11184546,11351331}. Specifically, the authors in \cite{10916605} investigated FA-enabled IDET networks and developed an analytical framework to characterize the joint information and energy transfer performance under different port selection strategies. By combining stochastic geometry and Copula theory, they showed that fluid antennas can provide significant performance gains over conventional static antennas in IDET systems. In \cite{10980171}, the authors studied a fluid antenna multiple access (FAMA)-assisted IDET system and derived analytical expressions for the outage probabilities and multiplexing gains under different port selection strategies. They further investigated a FAMA-enabled wireless powered communication network (WPCN) and analyzed the downlink and uplink outage performance in \cite{11184149}. In \cite{10930949}, the port activation and time scheduling were jointly optimized  to maximize the sum throughput of DRs in FA assisted WPCN. The authors in \cite{10506795} jointly optimized beamforming and port selection to maximize the weighted harvested energy in a FAS-assisted multiuser IDET system. Furthermore, \cite{11184546} 	first considered the antenna switching delay and the energy consumption caused by FA movement during the beamforming and port selection design. In \cite{11351331},the authors investigated IDET in multi-IRS-aided cell-free IoT networks with fluid antennas and proposed a joint optimization framework for FA port selection, transmit beamforming, and IRS reflection design to improve both information transmission and energy harvesting performance. 

Some works studied index modulation in FAS assisted systems \cite{10440042,11206363,10653737,11367092,11184847}. In \cite{10440042}, a fluid antenna index modulation scheme was proposed for MIMO systems, where the information bits are mapped to both symbols and antenna position patterns using an optimized codebook.  To mitigate the high spatial correlation among activated ports, \cite{11206363} proposed a fluid antenna group index modulation scheme. In \cite{10653737}, a single RF chain is shared by all ports in the fluid antenna system, and information transmission is realized by controlling the signal to select the target port group and the optimal port within that group, namely position index  modulation (PIM). Based on the signal-to-noise-ratio (SNR) maximization criterion in each port group, the authors in \cite{11367092} introduced PIM into traditional IRS assisted IDET system, where the transmit power is minimized by jointly optimizing the beamforming at BS and IRS. In \cite{11184847}, the authors proposed a FAS-assisted differential index modulation scheme for MIMO systems. In this paper, the term fluid index modulation (FIM) is used to refer to fluid antenna index modulation.

\subsection{Motivation and Contributions}
Although IDET, FAS, and IM have each been extensively studied, very limited attention has been paid to FIM-assisted IDET systems. In such a system, the integration of FAS and IM introduces both unique opportunities and significant challenges. On the one hand, the large number of available FA ports provides abundant spatial candidates for index modulation, offering additional degrees of freedom to improve both spectral efficiency and energy transfer performance. On the other hand, ports with strong channel gains are often highly spatially correlated, giving rise to a fundamental trade-off between energy harvesting and information decoding. Specifically, from an energy-transfer perspective, ports with stronger channel gains are preferred to maximize the harvested power. Conversely, from an information-transfer perspective, highly correlated ports complicate index detection and degrade the bit error rate (BER) performance. Therefore, reliable data transmission inherently favors ports with lower correlation and a larger minimum Euclidean distance, necessitating a sophisticated port selection strategy. Although \cite{11367092} introduced group-based FIM into a traditional IRS-assisted IDET system, it does not explicitly characterize the inherent trade-off between energy harvesting and information decoding. Moreover, reducing spatial correlation among selected ports via grouping is insufficient for compact FAS scenarios with densely distributed ports. In addition, it does not address port selection optimization under finite-alphabet signaling, which is critical for practical IM systems. 
Motivated by the above observations, this work studies an enhanced FIM scheme for IDET system with finite alphabet, where the port selection, precoding and power splitting ratio are jointly optimized to maximize the average harvested power while satisfying the BER and data rate constraints. 
The main contributions of this paper are summarized as follows:
\begin{itemize}
	\item We investigate an enhanced FIM-assisted IDET system, where the BS is equipped with a two-dimensional (2D) FAS and the IDET receiver adopts a power-splitting architecture. In the considered framework, information bits are conveyed jointly through the conventional amplitude-phase domain and the fluid index domain, enabling simultaneous WDT and WET.
	
	\item We theoretically analyze the average harvested power, BER, and achievable data rate under finite-alphabet signaling. Based on these results, a joint optimization problem is formulated to maximize the average harvested power by jointly designing the port selection, precoding vector, and power splitting ratio, while satisfying the BER and achievable rate constraints. To solve this highly coupled non-convex problem, we develop an alternating optimization (AO) algorithm that decomposes the original problem into two subproblems. In particular, the port selection subproblem is addressed by a block coordinate descent (BCD) method, while the precoding subproblem is solved by a Riemannian augmented Lagrangian method (RALM).
	
	\item Simulation results verify the effectiveness of the proposed enhanced FIM-assisted IDET design and demonstrate its superiority over several benchmark FIM schemes. In addition, the results explicitly characterize the R-E trade-off and show how the proposed design can effectively balance communication reliability and energy transfer performance.
\end{itemize}

The remainder of this paper is organized as follows. The enhanced FIM assisted IDET system model is introduced in Section \ref{section:architecture}. The average harvested power, BER and achievable data rate are analyzed in Section \ref{section:performance analysis}. Section \ref{section:problem solution} presents an AO algorithm for solving the optimization problem.  Simulation results are provided in Section \ref{section:sim}. Finally, Section \ref{section:conclusion} concludes this paper.

\textit{Notations:}
In this paper, matrices, vectors, and scalars are denoted by boldface uppercase letters, boldface lowercase letters, and lowercase letters, respectively. The sets of \(x\times y\) complex and real matrices are represented by \(\mathbb{C}^{x\times y}\) and \(\mathbb{R}^{x\times y}\), respectively. For a complex number \(x\), \(|x|\) denotes its modulus. For a vector \(\boldsymbol{x}\), \(\|\boldsymbol{x}\|\) denotes the Euclidean norm. The superscripts \((\cdot)^T\) and \((\cdot)^H\) denote the transpose and Hermitian transpose, respectively. Finally, \(\mathbb{E}[\cdot]\) denotes the expectation operator.

\section{System Model}\label{section:architecture}
\subsection{Transmit Signal Model}
\begin{figure*}
	\centering
	\includegraphics[width=1\linewidth]{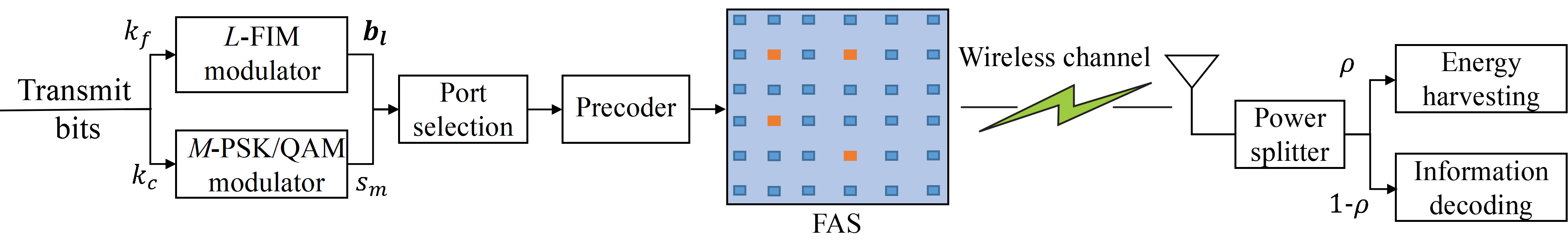}
	\caption{System model of FIM assisted IDET system.}
	\label{fig:system_model}
\end{figure*}

As shown in Fig. \ref{fig:system_model}, we consider a FIM assisted IDET system, where the BS is equipped with a two-dimensional fluid antenna (FA), while the IDET receiver employs a conventional fixed-position antenna due to hardware space limitations.  The FA at the BS has $N=N_1\times N_2$ ports uniformly distributed  over an area of $\lambda W_1\times \lambda W_2$, where $\lambda$ is the wavelength of RF signal,  $N_1$ and $N_2$ denote the number of ports along the vertical and horizontal directions, and $W_1$ and $ W_2$ are the corresponding scaling constants.  We assume that the FA is equipped with a single RF chain and can instantaneously switch to any of its ports. At the transmitter, $L$ out of $N$ ports are selected for the FIM transmission. A larger $L$ enhances the achievable data rate by increasing the number of index combinations, while it also results in stronger spatial correlation among the selected ports in FIM, which may degrade detection performance and even reduce the energy harvesting efficiency. Therefore, precoding is essential for each selected port to mitigate channel correlation and enhance demodulation performance. Since the FA is equipped with a single RF chain, only analog beamforming can be adopted at the transmitting FA. In this case, the precoding vector solely adjusts the phase while keeping the amplitude constant. The resultant precoding vector is denoted by $\mathbf{w}=[w_1,\cdots,w_L]\in\mathbb{C}^{L\times 1}$, satisfying $|w_l|=1$. Owing to the dense deployment of ports within a finite region, the channels associated with different FA ports exhibit strong correlation. Therefore, the ports selected for transmission and precoder need to be carefully designed. 

During each transmission, the transmit bits are divided into two parts, where $k_c=\log_2(M)$ bits are used for the classic $M$-ary PSK/QAM and $k_f=\log_2 (L)$ bits indicate the activated port.  Consequently, each transmitted symbol carries a total of $\log_2(ML)$ bits. The modulated symbol in conventional amplitude and phase modulation is denoted by $b_m$, satisfying $\mathbb{E}[|b_m|^2]=1$. In addition, the activation indicator $\mathbf{a}_l\in \mathbb{R}^{L\times 1}$ is employed to specify the activated port among the $L$ selected ports, with a single element equal to 1 and all remaining elements equal to 0. Therefore, the data symbol $\mathbf{s}_{m,l}=\mathbf{a}_lb_m$ is obtained. The transmitted signal can be expressed as 
\begin{align}
	\mathbf{x}_{m,l}=\sqrt{P_s}\text{diag}\left( \mathbf{w}\right) \mathbf{s}_{m,l}
\end{align}
where \(P_s\) denotes the transmit power, and \(\mathrm{diag}(\cdot)\) represents the diagonal matrix whose diagonal entries are given by the elements of the input vector.

\subsection{Wireless Channel Model}
We assume a block fading channel model between the transmitter and the receiver, in which the channel remains constant over a coherence time but changes independently from one frame to another. To characterize the spatial correlation among FA ports, a two-dimensional Cartesian coordinate system with the \(x\)- and \(y\)-axes is established on the plane where the FA is deployed. In this coordinate system, the position of the \(i\)-th port is denoted by \(\mathbf{t}_i = (x_i, y_i)\), where \(i \in \{1,2,\dots,N\}\). Assuming a rich-scattering three-dimensional propagation environment \cite{10103838}, the spatial correlation between the \(i\)-th and \(j\)-th ports can be expressed as
\begin{equation}
	J_{i,j} = {j}_0\!\left( 2\pi\sqrt{\left( \frac{|x_i-x_j|}{N_1-1}W_1\right)^2+\left(\frac{|y_i-y_j|}{N_2-1}W_2 \right)^2 } \right)
\end{equation}
where $j_0(\cdot)$ is the zero-order spherical Bessel function or the sinc function. Therefore, the correlation between any two ports can be represented by a matrix $\mathbf{J}=[J_{i,j}]$, which is a positive semidefinite matrix and can be decomposed as
$
	\mathbf{J}=\mathbf{U}\mathbf{\Lambda}\mathbf{U}^{H},
$
where \(\mathbf{U}\) denotes the unitary eigenvector matrix of \(\mathbf{J}\), and \(\mathbf{\Lambda}\) is a diagonal matrix whose diagonal entries are the eigenvalues of \(\mathbf{J}\). Accordingly, the complex full channels between the transmitter and the receiver can be modeled as
\begin{align}
	\mathbf{h}=\sqrt{\frac{1}{L_p}}\,\mathbf{U}\mathbf{\Lambda}^{1/2}\mathbf{g},
\end{align}
where \(\mathbf{g}\in\mathbb{C}^{N\times 1}\) is a complex Gaussian random vector whose entries are independently and identically distributed with zero mean and unit variance, and \(L_p\) denotes the path loss. Since \(L\) ports are selected from the total \(N\) ports for FIM, the channel corresponding to the selected ports can be expressed as
$
	\mathbf{h}_s=\mathbf{P}^T\mathbf{h},
$
where \(\mathbf{P}=[\mathbf{p}_1,\cdots,\mathbf{p}_L]\in\mathbb{R}^{N\times L}\) denotes the port selection matrix. Each selection vector \(\mathbf{p}_l\in\mathbb{R}^{N\times 1}\) contains only one entry equal to 1, while all the remaining entries are 0. The position of the nonzero entry indicates the index of the selected port. To ensure that the same port is not selected more than once, the selection vectors satisfy \(\mathbf{p}_l^T\mathbf{p}_k=0\), which guarantees that different selected ports are mutually exclusive.
\subsection{Receive Signal Model}
At the receiver side, IDET is realized by using a power splitting architecture. Specifically, the received RF signal is divided into two streams according to the power splitting ratio \(\rho\). A portion \(\sqrt{\rho}\) is allocated to energy harvesting, whereas the remaining portion \(\sqrt{1-\rho}\) is used for information demodulation. Based on this structure, the signal entering the energy harvesting branch is expressed as
\begin{align}
	{y}_{\mathrm{E}}&=\sqrt{\rho}\mathbf{h}_s^H\mathbf{x}_{m,l}+{z}\\ \notag
	&= \sqrt{\rho}\mathbf{h}^H\mathbf{P}\mathbf{x}_{m,l}+{z}
\end{align}
where \({z}\sim\mathcal{CN}(0,\sigma_z^2)\) is the additive white Gaussian noise (AWGN)  at the receive antenna. In addition, the received signal for information demodulation can be expressed as 
\begin{align}
	{y}_{\mathrm{I}}
	&= \sqrt{1-\rho}\mathbf{h}^H\mathbf{P}\mathbf{x}_{m,l}+{z}
\end{align}
Assuming that the receiver has perfect knowledge of the channel state information (CSI) of the selected ports, the maximum-likelihood (ML) detector is employed to recover the transmitted symbol by exhaustively searching over all possible transmit symbols. Therefore, the demodulation process can be formulated as
\begin{align}
(\overline{m},\overline{l})=arg \ \min_{\substack{m'\in[1,\cdots,M]\\l'\in[1,\cdots,L]}}\left(|{y}_{\mathrm{I}}-\sqrt{1-\rho}\mathbf{h}^H\mathbf{P}\mathbf{x}_{m',l'}|^2\right),
\end{align}
where $\overline{m}$ is the estimated index of the conventionally modulated symbol, $\overline{l}$ is the estimated index of transmit port, $m'$ and $l'$ are the trial index of conventionally modulated symbol and transmit port, respectively.

\section{WDT and WET Performance Analysis}\label{section:performance analysis}
\subsection{WET Performance}
In practical energy harvesting circuits, the rectification process is inherently nonlinear due to the characteristics of diodes. Therefore, a nonlinear energy harvesting model is adopted in this paper. By ignoring the noise power and adopting a commonly used polynomial-based nonlinear model \cite{8476597}, the average harvested power can be expressed as
\begin{align}
	E_r 
	&= \eta \left( k_2 \mathbb{E}\left[|\sqrt{\rho}\mathbf{h}^H\mathbf{P}\mathbf{x}_{m,l}|^2\right] 
	+ k_4 \mathbb{E}\left[|\sqrt{\rho}\mathbf{h}^H\mathbf{P}\mathbf{x}_{m,l}|^4\right] \right).
\end{align}
where  $\eta\in(0,1]$ represents the energy harvesting efficiency, $k_2$ and $k_4$ are circuit-dependent coefficients. In the considered system model, the transmit symbols  $\mathbf{x}_{m,l}$ are assumed to be equiprobable and have a unity average power. Therefore, the second-order term and the fourth-order term are given by
\begin{align}
	\mathbb{E}\left[|\mathbf{h}^H\mathbf{P}\mathbf{x}_{m,l}|^2\right]
	= \frac{P_s}{L}\sum_{l=1}^{L}|\mathbf{h}^H\mathbf{p}_l|^2,
\end{align}
\begin{align}
	\mathbb{E}\left[|\mathbf{h}^H\mathbf{P}\mathbf{x}_{m,l}|^4\right]
	= \frac{P_s^2}{ML}\sum_{m=1}^{M}|b_m|^4 \sum_{l=1}^{L}|\mathbf{h}^H\mathbf{p}_l|^4.
\end{align}
It is worth noting that for $M$-PSK or square $M$-QAM modulation, the fourth-order moment of the transmit symbols is a constant given by
\begin{equation}
\xi=\frac{1}{M}\sum_{m=1}^{M}|b_m|^4
 =
	\begin{cases}
		1, & \text{for } M\text{-PSK},\\[4pt]
		\dfrac{7M-13}{5(M-1)}, & \text{for square } M\text{-QAM}.
	\end{cases}
\end{equation}
Therefore, the average harvested power under the nonlinear model is given by
\begin{align}
	E_r 
	= \eta \left( k_2 \rho \frac{P_s}{L}\sum_{l=1}^{L}|\mathbf{h}^H\mathbf{p}_l|^2
	+ k_4 \rho^2 \frac{\xi P_s^2}{L} \sum_{l=1}^{L}|\mathbf{h}^H\mathbf{p}_l|^4 \right).\label{average_HP}
\end{align}
\subsection{WDT Performance}
In this section, the BER and achievable data rate performance of the proposed enhanced FIM assisted IDET system are analyzed based on the ML detection rule, which can be regarded as a lower-bound of any other demodulation approach. First, we definite the alphabetical set of the modulated symbols as $\mathcal{S}=\{\mathbf{s}_{m,l}|m\in\{1,\cdots,M\},l\in\{1,\cdots,L\}\}$ with cardinality $|\mathcal{S}|=ML$. Based on \cite{10491109}, BER can be expressed as
\begin{align}
	\epsilon_e=\frac{1}{k|\mathcal{S}|}\sum\limits_{\mathbf{s}_{m,l}\in\mathcal{S}}\sum\limits_{\substack{\mathbf{s}_{m',l'}\in\mathcal{S}\\\mathbf{s}_{m',l'}\neq \mathbf{s}_{m,l}}}d(\mathbf{s}_{m,l},\mathbf{s}_{m',l'})\Pr(\mathbf{s}_{m,l}\rightarrow \mathbf{s}_{m',l'})
\end{align}
where $k=k_c+k_f$ is the total number of bits carried by each symbol. \(d(\mathbf{s}_{m,l},\mathbf{s}_{m',l'})\) denotes the Hamming distance between the information bits mapped to \(\mathbf{s}_{m,l}\) and \(\mathbf{s}_{m',l'}\). Moreover, \(\Pr(\mathbf{s}_{m,l}\rightarrow \mathbf{s}_{m',l'})\) represents the probability that the transmitted symbol \(\mathbf{s}_{m,l}\) is erroneously detected as \(\mathbf{s}_{m',l'}\). Therefore, \(\Pr(\mathbf{s}_{m,l}\rightarrow \mathbf{s}_{m',l'})\) can be derived as 
\begin{align}\label{P_ect}
	\Pr(\mathbf{s}_{m,l}\rightarrow \mathbf{s}_{m',l'})&=\Pr	\left[|y_I-\sqrt{1-\rho}\mathbf{h}^H\mathbf{P}\mathbf{x}_{m',l'}|^2\right.\notag\\
	&\left.<\min_{\mathbf{x}_{n,i}\neq\mathbf{x}_{m',l'}}|y_I-\sqrt{1-\rho}\mathbf{h}^H\mathbf{P}\mathbf{x}_{n,i}|^2\right]. 
\end{align}
Nevertheless, obtaining a closed-form expression for \eqref{P_ect} is challenging, because the term 
\(|y_I-\sqrt{1-\rho}\mathbf{h}^H\mathbf{P}\mathbf{x}_{n,i}|^2\) 
is statistically dependent on the corresponding terms associated with other candidates \(\mathbf{x}_{m',l'}\). Therefore, we resort to an asymptotic analysis in the high SNR regime. In this case, the Euclidean distance between the received signal \(y_I\) and the transmitted signal component \(\sqrt{1-\rho}\mathbf{h}^H\mathbf{P}\mathbf{x}_{m,l}\) is most likely to be the smallest among all candidate symbols.
Hence, we have the following approximation
\begin{align}
	\min_{\substack{\mathbf{x}_{n,i}\neq \\
			 \mathbf{x}_{m',l'}}}
	\left|y_I-\sqrt{1-\rho}\mathbf{h}^H\mathbf{P}\mathbf{x}_{n,i}\right|^2
	\approx
	\left|y_I-\sqrt{1-\rho}\mathbf{h}^H\mathbf{P}\mathbf{x}_{m,l}\right|^2
\end{align}
and 
\begin{align} \small
\Pr(\mathbf{s}_{m,l}\rightarrow \mathbf{s}_{m',l'})&\approx \Pr	\left[|y_I-\sqrt{1-\rho}\mathbf{h}^H\mathbf{P}\mathbf{x}_{m',l'}|^2\right.\notag\\
&\left.<|y_I-\sqrt{1-\rho}\mathbf{h}^H\mathbf{P}\mathbf{x}_{m,l}|^2\right].
\end{align}
 
By defining the pairwise error probability (PEP) as $\tau(\mathbf{s}_{m,l}\rightarrow \mathbf{s}_{m',l'})=\Pr	\left[|y_I-\sqrt{1-\rho}\mathbf{h}^H\mathbf{P}\mathbf{x}_{m',l'}|^2<|y_I-\sqrt{1-\rho}\mathbf{h}^H\mathbf{P}\mathbf{x}_{m,l}|^2\right]$, the upper bound of BER can be obtained by 
\begin{align}
	\epsilon=\frac{1}{k|\mathcal{S}|}\sum\limits_{\mathbf{s}_{m,l}\in\mathcal{S}}\sum\limits_{\substack{\mathbf{s}_{m',l'}\in\mathcal{S}\\\mathbf{s}_{m',l'}\neq \mathbf{s}_{m,l}}}d(\mathbf{s}_{m,l},\mathbf{s}_{m',l'})\tau(\mathbf{s}_{m,l}\rightarrow \mathbf{s}_{m',l'}),
\end{align}
where 
\begin{align}
	\tau(\mathbf{s}_{m,l}&\rightarrow \mathbf{s}_{m',l'})=Q\left( \sqrt{\frac{(1-\rho)|\mathbf{h}^H\mathbf{P}(\mathbf{x}_{m,l}-\mathbf{x}_{m',l'})|^2}{2\sigma_z^2}}\right) \notag \\ 
	&=Q\left( \sqrt{\frac{(1-\rho)P_s|\mathbf{h}^H\mathbf{p}_lw_lb_m-\mathbf{h}^H\mathbf{p}_{l'}w_{l'}b_{m'}|^2}{2\sigma_z^2}}\right).
\end{align}
$Q(x)$ is the Q-function given by
$
	Q(x)=\int_x^{+\infty}\frac{1}{\sqrt{2\pi}}e^{-\frac{t^2}{2}}\,dt.
$
The upper bound BER $\epsilon$ is used in the subsequent achievable data rate analysis and optimization. 

Based on the BER expression, the achievable data rate of the proposed system can be evaluated according to \cite{10491109}. Specifically, the end-to-end link between the transmitter and the receiver is modeled as a memoryless binary symmetric channel (BSC) with crossover probability \(\epsilon\). Therefore, the resultant achievable data rate is given by
\begin{align}
	R
	=
	k I(x;\hat{y})
	=
	k\left[
	1
	+
	\epsilon\log_2\epsilon
	+
	(1-\epsilon)\log_2(1-\epsilon)
	\right].
\end{align}

\section{Joint Port Selection and Precoding Design}\label{section:problem solution}
\subsection{Problem Formulation}
Owing to the higher spatial degrees of freedom provided by FAS, FAS assisted IDET systems can potentially achieve superior performance in both information delivery and energy transfer. However, in the proposed enhanced FIM assisted IDET systems, the performances of these two functions are strongly coupled with the port selection strategy. Specifically, for reliable information transmission, the selected ports are expected to maximize the minimum Euclidean distance so as to reduce the BER. In contrast, for efficient energy transfer, the selected ports are preferably those associated with larger channel magnitudes. Unfortunately, in FAS, ports with large channel magnitudes are usually highly correlated, which makes port index information detection more challenging. Motivated by this observation, a precoding vector is introduced to reduce the correlation among the selected ports, with the aim of enhancing BER performance without sacrificing energy transfer performance. Therefore, in this section, we jointly optimize the port selection, precoding vector, and power splitting ratio to maximize the average harvested power under BER and achievable rate constraints, in order to characterize the trade-off between WDT and WET. The optimization problem can be formulated as
 \begin{align}
	\text{(P1)}:\max_{\{\mathbf{p}_l\},\mathbf{w},\rho} \quad & E_r \label{obj0}\\
	s.t.  \ \ \quad &\epsilon\leq \epsilon_{th}, \tag{\ref{obj0}{a}}\label{obj0a}\\
	&R\ge R_{th}, \tag{\ref{obj0}{b}}\label{obj0b}\\
	&\mathbf{p}_l^T\mathbf{p}_k=0, \forall l\neq k \tag{\ref{obj0}{c}}\label{obj0c}\\
	&\mathbf{p}_l^T\mathbf{1}^{N}=1, \forall l=1,\cdots,L \tag{\ref{obj0}{d}}\label{obj0d}\\
	&\mathbf{p}_l\in\{0,1\}^{N},\tag{\ref{obj0}{e}}\label{obj0e}\\
	&|w_l|=1, \forall l=1,\cdots,L \tag{\ref{obj0}{f}}\label{obj0f}\\
	&0<\rho< 1 \tag{\ref{obj0}{g}}\label{obj0g},
\end{align}
where \(\epsilon_{th}\) and \(R_{th}\) denote the BER threshold and the achievable rate threshold, respectively. It is worth noting that the achievable rate \(R\) is monotonically decreasing with respect to \(\epsilon\) when \(\epsilon<0.5\). Therefore, the rate constraint can be equivalently transformed into \(\epsilon \leq \epsilon_{R,th}\), where \(\epsilon_{R,th}\) satisfies
$
	R_{th}=k\left[1+\epsilon_{R,th}\log_2\epsilon_{R,th}+(1-\epsilon_{R,th})\log_2(1-\epsilon_{R,th})\right].
$
As a result, \eqref{obj0a} and \eqref{obj0b} can be merged into a single constraint, i.e.,
$
\epsilon \leq \epsilon_{th}',
$
where $\epsilon_{th}'=\min\{\epsilon_{th},\epsilon_{R,th}\}$. Accordingly, problem (P1) can be reformulated as
 \begin{align}
	\text{(P2)}:\max_{\{\mathbf{p}_l\},\mathbf{w},\rho} \quad & E_r \label{obj1}\\
	s.t.  \ \ \quad &\epsilon\leq \epsilon_{th}', \tag{\ref{obj1}{a}}\label{obj1a}\\
	&\mathbf{p}_l^T\mathbf{p}_k=0, \forall l\neq k \tag{\ref{obj1}{b}}\label{obj1b}\\
	&\mathbf{p}_l^T\mathbf{1}^{N}=1,\forall l=1,\cdots,L, \tag{\ref{obj1}{c}}\label{obj1c}\\
	&\mathbf{p}_l\in\{0,1\}^{N},\tag{\ref{obj1}{d}}\label{obj1d}\\
	&|w_l|=1, \forall l=1,\cdots,L,\tag{\ref{obj1}{e}}\label{obj1e}\\
	&0<\rho< 1 \tag{\ref{obj1}{f}}\label{obj1f}.
\end{align}
 Note that (P2) is a non-convex mixed-integer programming problem and is generally NP-hard, which is hard to solve optimally in a direct manner. To address this issue, we first reformulate (P2) into a more tractable form and further decompose it into two relatively simple subproblems. Then, an alternating optimization (AO) algorithm is employed to iteratively solve the subproblems. This treatment significantly simplifies the solution procedure of the original problem and improves the computational efficiency.

\subsection{Problem Transformation and Solution}
\subsubsection{Problem transformation}
The BER constraint is non-convex due to the presence of a sum of \(Q\)-functions. To facilitate tractable optimization, we first relax it into an upper-bound form. Note that  $Q(x)$ is monotonically decreasing in $x$, thus we have   
\begin{align}
	&\tau(\mathbf{s}_{m,l}\rightarrow \mathbf{s}_{m',l'})\notag\\
	&=Q\left( \sqrt{\frac{(1-\rho)P_s|\mathbf{h}^H\mathbf{p}_lw_lb_m-\mathbf{h}^H\mathbf{p}_{l'}w_{l'}b_{m'}|^2}{2\sigma_z^2}}\right)\notag\\
	&\leq Q\left( \sqrt{\frac{(1-\rho)P_sd_{\min}(\{\mathbf{p}_l\},\mathbf{w})}{2\sigma_z^2}}\right)=\epsilon_0
\end{align}
where 
\begin{align}
	d_{\min}(\{\mathbf{p}_l\},\mathbf{w}) = \min_{(m,l)\neq (m',l')} 
	\left| \mathbf{h}^H \mathbf{p}_l w_l b_m - \mathbf{h}^H \mathbf{p}_{l'} w_{l'} b_{m'} \right|^2
\end{align}
denotes the minimum squared Euclidean distance. The BER constraint in \eqref{obj1a} can then be expressed as
\begin{align}
	\epsilon&=\frac{1}{kML}\sum\limits_{\mathbf{s}_{m,l}\in\mathcal{S}}\sum\limits_{\substack{\mathbf{s}_{m',l'}\in\mathcal{S}\\\mathbf{s}_{m',l'}\neq \mathbf{s}_{m,l}}}d(\mathbf{s}_{m,l},\mathbf{s}_{m',l'})\tau(\mathbf{s}_{m,l}\rightarrow \mathbf{s}_{m',l'})\notag\\
	&\leq\frac{1}{kML}\sum\limits_{\mathbf{s}_{m,l}\in\mathcal{S}}\sum\limits_{\substack{\mathbf{s}_{m',l'}\in\mathcal{S}\\\mathbf{s}_{m',l'}\neq \mathbf{s}_{m,l}}}d(\mathbf{s}_{m,l},\mathbf{s}_{m',l'})\epsilon_0\leq\epsilon_{th}'.
\end{align}
Thus we can obtain 
\begin{align}\label{qq}
	\epsilon_0&=Q\left( \sqrt{\frac{(1-\rho)P_sd_{\min}(\mathbf{p}_l,\mathbf{w})}{2\sigma_z^2}}\right)\notag\\
	&\leq\frac{kML\epsilon_{th}'}{\sum\limits_{\mathbf{s}_{m,l}\in\mathcal{S}}\sum\limits_{\substack{\mathbf{s}_{m',l'}\in\mathcal{S}\\\mathbf{s}_{m',l'}\neq \mathbf{s}_{m,l}}}d(\mathbf{s}_{m,l},\mathbf{s}_{m',l'})}=\gamma_{th}.
\end{align}
Since $Q(x)$ is monotonically decreasing, the inequality in Eq. \eqref{qq} can be equivalently rewritten as 
\begin{align}
	(1-\rho)d_{\min}(\{\mathbf{p}_l\},\mathbf{w})\ge \frac{2\sigma_z^2(Q^{-1}(\gamma_{th}))^2}{P_s}.
\end{align}
Therefore, by replacing the BER constraint with the sufficient
minimum-distance constraint, which tightens the feasible region, (P2) can be conservatively reformulated as
\begin{align}
	\text{(P3)}:\max_{\{\mathbf{p}_l\},\mathbf{w},\rho} \quad
	& E_r(\{\mathbf{p}_l\},\rho) \label{obj3}\\
	\text{s.t.}\quad
	& (1-\rho)d_{\min}(\{\mathbf{p}_l\},\mathbf{w}) \ge C,
	\tag{\ref{obj3}{a}}\label{obj3a}\\
	& \mathbf{p}_l^T\mathbf{p}_k=0,\ \forall l\neq k,
	\tag{\ref{obj3}{b}}\label{obj3b}\\
	& \mathbf{p}_l^T\mathbf{1}^{N}=1,\ \forall l=1,\cdots,L,
	\tag{\ref{obj3}{c}}\label{obj3c}\\
	& \mathbf{p}_l\in\{0,1\}^{N},
	\tag{\ref{obj3}{d}}\label{obj3d}\\
	& |w_l|=1,\ \forall l=1,\cdots,L,
	\tag{\ref{obj3}{e}}\label{obj3e}\\
	& 0<\rho<1,
	\tag{\ref{obj3}{f}}\label{obj3f}
\end{align}
where
$
C=\frac{2\sigma_z^2\left(Q^{-1}(\gamma_{th})\right)^2}{P_s}.
$
Therefore, the optimal value of (P3) serves as a lower bound on that of (P2). 
Due to the strong coupling among \(\{\mathbf{p}_l\}\), \(\mathbf{w}\), and
\(\rho\), problem (P3) is still non-convex.
To facilitate the solution, we first derive the optimal power splitting
ratio for any given \(\{\mathbf{p}_l\}\) and \(\mathbf{w}\).

For fixed \(\{\mathbf{p}_l\}\) and \(\mathbf{w}\), the harvested power
\(E_r\) is monotonically increasing with respect to \(\rho\). Specifically,
we have
\begin{align}
	\frac{\partial E_r}{\partial \rho}
	=
	\eta\left(
	\frac{k_2P_s}{L}S_2
	+
	\frac{2k_4\xi P_s^2}{L}\rho S_4
	\right) > 0,
\end{align}
where
$
S_2=\sum_{l=1}^{L}|\mathbf{h}^H\mathbf{p}_l|^2,
S_4=\sum_{l=1}^{L}|\mathbf{h}^H\mathbf{p}_l|^4.
$
Therefore, the optimal \(\rho\) is obtained when constraint
\eqref{obj3a} holds with equality, yielding
\begin{align}
	\rho^\star
	=
	1-\frac{C}{d_{\min}(\{\mathbf{p}_l\},\mathbf{w})}.
	\label{rho_close}
\end{align}
It should be noted that \(\rho^\star\) is feasible only when
\(d_{\min}(\{\mathbf{p}_l\},\mathbf{w})>C\). Otherwise, no feasible power
splitting ratio exists for the given \(\{\mathbf{p}_l\}\) and \(\mathbf{w}\).
Substituting \(\rho^\star\) into \(E_r\), the reduced harvested power can be
written as
\begin{align}
	\widetilde{E}_r(\{\mathbf{p}_l\},\mathbf{w})
	=
	\eta\left[
	\frac{k_2P_s}{L}\rho^\star S_2
	+
	\frac{k_4\xi P_s^2}{L}(\rho^\star)^2 S_4
	\right].
\end{align}
Then, (P3) can be solved by alternately optimizing the phase-shift vector
\(\mathbf{w}\) and the port selection vectors \(\{\mathbf{p}_l\}\). 
\subsubsection{Phase optimization on a manifold}

Given the port selection vectors \(\{\mathbf{p}_l\}\), both \(S_2\) and
\(S_4\) are independent of \(\mathbf{w}\). Moreover,
\(\widetilde{E}_r\) is monotonically increasing with
\(d_{\min}(\{\mathbf{p}_l\},\mathbf{w})\) when
\(d_{\min}(\{\mathbf{p}_l\},\mathbf{w})>C\). Therefore, the phase
optimization subproblem can be equivalently written as
\begin{align}
	\text{(P4)}:\max_{\mathbf{w}} \quad
	& d_{\min}(\mathbf{w}) \label{obj4}\\
	\text{s.t.}\quad
	& |w_l|=1,\ \forall l=1,\cdots,L.
	\tag{\ref{obj4}{a}}\label{obj4a}
\end{align}
By introducing an auxiliary variable \(t\), (P4) can be rewritten in
epigraph form as
\begin{align}
	\text{(P5)}:\max_{\mathbf{w},t} \quad
	& t \label{obj5}\\
	\text{s.t.}\quad
	& \left|
	\mathbf{h}^H\mathbf{p}_l w_l b_m
	-
	\mathbf{h}^H\mathbf{p}_{l'}w_{l'}b_{m'}
	\right|^2 \ge t,
	\notag\\
	& \forall (m,l)\neq(m',l'),
	\tag{\ref{obj5}{a}}\label{obj5a}\\
	& |w_l|=1,\ \forall l=1,\cdots,L.
	\tag{\ref{obj5}{b}}\label{obj5b}
\end{align} 
It is worth noting that the number of constraints in \eqref{obj5a} is \(ML(ML-1)/2\), which is prohibitively large when high-order modulation is adopted. In addition, these constraints are all non-convex quadratic inequality constraints. Together with the unit-modulus constraints in \eqref{obj5b}, problem (P5) remains highly non-convex and thus is difficult to solve directly.

To address this issue, we employ a Riemannian augmented Lagrangian optimization framework \cite{deng2025}. Specifically, the unit-modulus constraints indicate that the feasible set of \(\mathbf{w}=[w_1,\cdots,w_L]^T\) lies on a complex circle manifold (CCM), defined as $\mathcal{M}=\{\mathbf{w}\in\mathbb{C}^L:|w_l|=1,\forall l\}$. Meanwhile, the Euclidean-distance constraints are incorporated into the objective function through an augmented penalty mechanism. In this way, the original constrained problem can be transformed into an unconstrained optimization problem defined on the manifold $\mathcal{M}$. To further streamline the expression, we map the dual indices \((m,l)\) to a single index \(i \in \{1,\ldots,ML\}\) using a bijective mapping \(\mathcal{I}(m,l)=(l-1)M+m\). Accordingly, let \(d_i=h_l b_m\) denote the effective channel-symbol product for the \(i\)-th combination. The distance constraint in \eqref{obj5a} can be rewritten as $g_{i,j}(\mathbf{w},t)=t-|d_iw_{l_i}-d_jw_{l_j}|^2\le 0$ for $i\neq j$
. The augmented Lagrangian function of (P5) can then be expressed as 
\begin{align}
	\mathcal{L}_{\beta}(\mathbf{w}, t, \boldsymbol{\alpha}) = -t + \frac{1}{2\beta} \sum_{i=1}^{ML} \sum_{j=i+1}^{ML} \left( \left[\alpha_{i,j} + \beta g_{i,j}(\mathbf{w}, t) \right]_{+}^2 - \alpha_{i,j}^2 \right) \label{AL_func_mapped}
\end{align}
where $[\cdot]_+ = \max(0, \cdot)$ ensures the non-negativity, $\beta>0$ is a penalty parameter, \(\boldsymbol{\alpha}=\{\alpha_{i,j}\}\) denotes Lagrange multiplier vector. Based on the augmented Lagrangian function in \eqref{AL_func_mapped}, problem (P5) can be solved by alternately optimizing  \(\mathbf{w},t\) for fixed \(\boldsymbol{\alpha}\) using a manifold optimization method, and updating \(\boldsymbol{\alpha}\) via a gradient-based rule. The overall optimization procedure follows a two-layer iterative structure. In the inner loop, for given multipliers \(\boldsymbol{\alpha}\)  and penalty parameter \(\beta\), the augmented Lagrangian function is minimized with respect to \((\mathbf{w},t)\). Since \(\mathbf{w}\) is constrained on the complex circle manifold and \(t\) lies in the Euclidean space \(\mathbb{R}\), the inner subproblem is defined on the product manifold $\mathcal{M}\times \mathbb{R}$, which can be solved by the Riemannian Conjugate Gradient (RCG) algorithm \cite{10737380}. Once the inner loop converges to a local optimum, the outer loop is triggered to update the Lagrange multipliers $\boldsymbol{\alpha}$ and monotonically increase the penalty parameter $\beta$ to heavily penalize the constraint violations. This alternating process repeats until the overall convergence criteria are met.

To execute the RCG algorithm within the inner loop, we must tailor the update rules to the distinct geometric properties of the product manifold $\mathcal{M} \times \mathbb{R}$. 
Since the auxiliary variable $t$ lies in the Euclidean space, its update follows the standard steepest descent direction. By defining an effective multiplier $\mu_{i,j} = \left[\alpha_{i,j} + \beta g_{i,j}(\mathbf{w}, t) \right]_{+}$, the Euclidean partial derivative of $\mathcal{L}_{\beta}$ with respect to $t$ is derived as
\begin{align}
	\nabla_t \mathcal{L}_{\beta}
	=
	-1+\sum_{i=1}^{ML}\sum_{j=i+1}^{ML}\mu_{i,j},
	\label{grad_t_final}
\end{align}
 Accordingly, the search direction for $t$ at the $r$-th inner iteration is set as $d_{t_r} = -\nabla_t \mathcal{L}_{\beta}$, and the subsequent point is updated via $t_{r+1} = t_r + \delta_r d_{t_r}$, where $\delta_r$ is the step size.

To handle the unit-modulus constraints of \(\mathbf{w}\), we employ the RCG algorithm. The search space is locally approximated by the tangent space at the current iterate \(\mathbf{w}_r \in \mathcal{M}\). Geometrically, this space consists of all complex vectors orthogonal to \(\mathbf{w}_r\), which is defined as
\begin{align}
	T_{\mathbf{w}_r}\mathcal{M}
	=
	\left\{
	\mathbf{z}\in\mathbb{C}^L
	\;\middle|\;
	\mathrm{Re}\left\{\mathbf{z}\odot \mathbf{w}_r^*\right\}
	=
	\mathbf{0}_L
	\right\},
\end{align}
where \(\odot\) denotes the Hadamard product. To determine the steepest descent direction that conforms to the manifold geometry, the Euclidean gradient \(\nabla_{\mathbf{w}_r}\mathcal{L}_{\beta}\) is orthogonally projected onto the tangent space \(T_{\mathbf{w}_r}\mathcal{M}\). The resulting Riemannian gradient is given by
\begin{align}
	\mathrm{grad}_{\mathbf{w}_r}\mathcal{L}_{\beta}
	=
	\nabla_{\mathbf{w}_r}\mathcal{L}_{\beta}
	-
	\mathrm{Re}\left\{
	\nabla_{\mathbf{w}_r}\mathcal{L}_{\beta}\odot \mathbf{w}_r^*
	\right\}
	\odot \mathbf{w}_r.
\end{align}
The Euclidean gradient \(\nabla_{\mathbf{w}_r}\mathcal{L}_{\beta}\)  is given by 
\begin{align}
	\nabla_{\mathbf{w}_r}\mathcal{L}_{\beta}= \sum_{i=1}^{ML}\sum_{j=i+1}^{ML}\mu_{i,j}(d_i^*d_jw_{l_j}\mathbf{e}_{l_i}+d_id_j^*w_{l_i}\mathbf{e}_{l_j}),
\end{align}
where \(\mathbf{e}_{l}\in\mathbb{R}^{L}\) denotes the \(l\)-th canonical basis vector.
Unlike the standard steepest descent method, RCG accelerates convergence by incorporating historical gradient information. However, since the tangent spaces at \(\mathbf{w}_{r-1}\) and \(\mathbf{w}_r\) are different, vectors in these two tangent spaces cannot be directly added. To address this issue, a vector transport operator \(\mathcal{T}_{\mathbf{w}_{r-1}\to \mathbf{w}_r}(\cdot)\) is introduced to map the previous search direction \(\mathbf{d}_{\mathbf{w}_{r-1}}\) to the current tangent space. The transport operator is defined as
\begin{align}
	\mathcal{T}_{\mathbf{w}_{r-1}\to \mathbf{w}_r}(\mathbf{d}_{\mathbf{w}_{r-1}})
	=
	\mathbf{d}_{\mathbf{w}_{r-1}}
	-
	\mathrm{Re}\left\{
	\mathbf{d}_{\mathbf{w}_{r-1}}\odot \mathbf{w}_r^*
	\right\}
	\odot \mathbf{w}_r,
	\label{transport_op}
\end{align}
Consequently, the conjugate search direction for \(\mathbf{w}\) is generated as
\begin{align}
	\mathbf{d}_{\mathbf{w}_r}
	=
	-\mathrm{grad}_{\mathbf{w}_r}\mathcal{L}_{\beta}
	+
	\Gamma_r\,
	\mathcal{T}_{\mathbf{w}_{r-1}\to \mathbf{w}_r}
	\big(\mathbf{d}_{\mathbf{w}_{r-1}}\big),\label{csd}
\end{align}
where \(\Gamma_r\) is the conjugate parameter updated according to the Polak--Ribière (PR+) rule to accelerate the convergence rate \cite{11175171}. It is given by
\begin{align}
	\Gamma_r
	=\left[ 
	\frac{
		\mathrm{Re}\left\lbrace 
		\mathrm{grad}_{\mathbf{w}_r}^H\mathcal{L}_{\beta}
		\left( \mathrm{grad}_{\mathbf{w}_r}\mathcal{L}_{\beta}
	-
		\mathrm{grad}_{\mathbf{w}_{r-1}}\mathcal{L}_{\beta}\right) 
		\right\rbrace 
	}{	
		||\mathrm{grad}_{\mathbf{w}_{r-1}}\mathcal{L}_{\beta}||^2
	}\right]_+ .
\end{align}

Finally, moving along the tangent direction \(\mathbf{d}_{\mathbf{w}_r}\) generally causes the phase-shift vector to leave the manifold. Therefore, a retraction mapping is required to pull the updated point back onto \(\mathcal{M}\). For the complex circle manifold, this operation is implemented by element-wise normalization:
\begin{align}
	\mathbf{w}_{r+1}
	=\text{unt}(\mathbf{w}_r + \delta_r \mathbf{d}_{\mathbf{w},r})=
	\frac{\mathbf{w}_r+\delta_r\mathbf{d}_{\mathbf{w}_r}}
	{\left|\mathbf{w}_r+\delta_r\mathbf{d}_{\mathbf{w}_r}\right|}.
\end{align}
where $\text{unt}(x)=x/|x|$ and the step size $\delta_r > 0$ is determined by using the Armijo backtracking line search to guarantee the sufficient decrease of the augmented Lagrangian function \cite{10660518}. With these operations, the inner loop proceeds iteratively to search for a critical point where the Riemannian gradient of the augmented Lagrangian function $\mathcal{L}_{\beta}$ vanishes. By employing the RCG method on the product manifold $\mathcal{M} \times \mathbb{R}$, the algorithm effectively exploits the coupling between the phase-shift vector $\mathbf{w}$ and the auxiliary variable $t$, leading to a more robust descent compared to standard first-order methods.

Upon reaching the inner convergence criterion, the optimized pair $(\mathbf{w}_{r+1}, t_{r+1})$ is passed to the outer loop to update the Lagrange multipliers $\boldsymbol{\alpha}$ and the penalty parameter $\beta$. Specifically, the multipliers are updated based on the degree of constraint violation at the current local optimum to gradually enforce feasibility. Let $k$ denote the index of the outer iteration, and let $(\mathbf{w}^{(k+1)}, t^{(k+1)})$ be the converged solution obtained from the $k$-th inner loop. The Lagrange multipliers are updated as follows:
\begin{align}
	\alpha_{i,j}^{(k+1)} = \text{clip}_{[\alpha_{\min}, \alpha_{\max}]} \left( \alpha_{i,j}^{(k)} + \beta^{(k)} g_{i,j}\big(\mathbf{w}^{(k+1)}, t^{(k+1)}\big) \right), \label{multiplier_clip_update}
\end{align}
where the clipping function is used to confine each multiplier within a predefined range to ensure numerical stability during the manifold optimization.  Instead of monotonically increasing \(\beta_k\) at every outer iteration, we adopt an adaptive update rule. Let
\begin{align}
	v_k=\max_{i<j}\left[g_{i,j}(\mathbf{w}^{(k)},t^{(k)})\right]_+ .
\end{align}
Then, \(\beta_k\) is updated as
\begin{align}
	\beta_{k+1}
	=
	\begin{cases}
		\beta_k, & \text{if } v_{k+1}\le \tau v_k,\\
		\zeta \beta_k, & \text{otherwise},
	\end{cases}
\end{align}
where \(0<\tau<1\) and \(\zeta>1\). This strategy prevents the penalty parameter from growing too fast when the constraints are already well improved. The procedure for solving (P5), named RALM, is summarized in Algorithm \ref{alg:RALM_RCG}. 

\begin{algorithm}[t]
	\renewcommand{\algorithmicrequire}{\textbf{Input:}}
	\renewcommand{\algorithmicensure}{\textbf{Output:}}
	\caption{RALM for solving (P5)}
	\label{alg:RALM_RCG}
	\begin{algorithmic}[1]
		\REQUIRE Initial point \(\mathbf{w}^{0}\in\mathcal{M}\), \(t^{0}\in\mathbb{R}\), multipliers \(\boldsymbol{\alpha}^{0}=\mathbf{0}\), penalty parameter \(\beta^{0}>0\); tolerances \(\varepsilon_{\mathrm{out}}\) and \(\varepsilon_{\mathrm{in}}\); maximum outer and inner iterations \(K_{\max}\) and \(R_{\max}\); ratio $\tau\in(0,1)$ and penalty scaling factor \(\zeta>1\).
		\ENSURE \(\mathbf{w}^\star,t^\star\).
		
		\FOR{$k=0,1,\ldots,K_{\max}$}
		\STATE Initialize inner variables: \(\mathbf{w}_{0}=\mathbf{w}^{k}\), \(t_{0}=t^{k}\).
		\STATE Compute the initial search directions:
		\[
		\mathbf{d}_{\mathbf{w},0}=-\mathrm{grad}_{\mathbf{w}_{0}}\mathcal{L}_{\beta},\qquad
		d_{t,0}=-\nabla_{t}\mathcal{L}_{\beta}.
		\]
		
		\FOR{$r=0,1,\ldots,R_{\max}$}
		\STATE Determine the step size \(\delta_{r}\) by Armijo backtracking line search according to \cite{10660518}.
		\STATE Update $t_{r+1}$: \(t_{r+1}=t_{r}+\delta_{r}d_{t,r}.\)
	
		\STATE Update $\mathbf{w}_{r+1}$ via retraction: $\mathbf{w}_{r+1} = \text{unt}(\mathbf{w}_r + \delta_r \mathbf{d}_{\mathbf{w},r})$
	
		\IF{$\|\mathrm{grad}\,\mathcal{L}_{\beta}(\mathbf{w}_{r+1},t_{r+1},\boldsymbol{\alpha}^{k})\| \le \varepsilon_{\mathrm{in}}$}
		\STATE \textbf{break}
		\ENDIF		
		\STATE Compute PR parameter \(\Gamma_{r+1}\) and update search direction $\mathbf{d}_{\mathbf{w}_{r+1}}$ according to \eqref{csd}.
		\ENDFOR
		
		\STATE Set	$\mathbf{w}^{k+1}=\mathbf{w}_{r+1},
		t^{k+1}=t_{r+1}.$
		
		\STATE Update the multipliers according to \eqref{multiplier_clip_update}.
\STATE Compute the maximum constraint violation
\[
v_{k+1}=\max_{i<j}\left[g_{i,j}(\mathbf{w}^{k+1},t^{k+1})\right]_+.
\]
\IF{$v_{k+1}\le \varepsilon_{\mathrm{out}}$ \textbf{and} \(\|\mathbf{w}^{k+1}-\mathbf{w}^{k}\|\le \varepsilon_{\mathrm{out}}\) \textbf{and} \(|t^{k+1}-t^{k}|\le \varepsilon_{\mathrm{out}}\)}
\STATE \textbf{break}
\ENDIF
\IF{$k=0$ or $v_{k+1}\le \tau v_k$}
\STATE Set \(\beta^{k+1}=\beta^k\).
\ELSE
\STATE Set \(\beta^{k+1}=\zeta \beta^k\).
\ENDIF
	
		\ENDFOR
		
		\STATE \textbf{return} \(\mathbf{w}^{\star}=\mathbf{w}^{k+1}\), \(t^{\star}=t^{k+1}\).
	\end{algorithmic}
\end{algorithm}
\subsubsection{Port selection optimization via BCD}

Given the optimized phase shift vector \(\mathbf{w}\), the port selection
vectors \(\{\mathbf{p}_l\}\) are optimized based on the reduced harvested
power \(\widetilde{E}_r\). Specifically, by substituting the closed-form
power splitting ratio \(\rho^\star\) into the nonlinear EH model, the port
selection subproblem can be written as
\begin{align}
	\text{(P6)}:\max_{\{\mathbf{p}_l\}} \quad
	& \widetilde{E}_r(\{\mathbf{p}_l\},\mathbf{w}) \label{obj6}\\
	\text{s.t.}\quad
	& d_{\min}(\{\mathbf{p}_l\},\mathbf{w})>C,
	\tag{\ref{obj6}{a}}\label{obj6a}\\
	& \mathbf{p}_l^T\mathbf{p}_k=0,\ \forall l\neq k,
	\tag{\ref{obj6}{b}}\label{obj6b}\\
	& \mathbf{p}_l^T\mathbf{1}^{N}=1,\ \forall l=1,\cdots,L,
	\tag{\ref{obj6}{c}}\label{obj6c}\\
	& \mathbf{p}_l\in\{0,1\}^{N}.
	\tag{\ref{obj6}{d}}\label{obj6d}
\end{align}
Problem (P6) is still difficult to solve globally due to the coupled binary
selection variables and the minimum-distance term. To address this issue,
we adopt a BCD method, where one port selection
vector is updated at a time while the remaining \(L-1\) selection vectors
are fixed.

In the update of the \(l\)-th block, the feasible candidate set is given by
\begin{align}
	\mathcal{N}_l
	=
	\{1,2,\ldots,N\}
	\setminus
	\bigcup_{k\neq l}\mathrm{supp}(\mathbf{p}_k),
\end{align}
where \(\mathrm{supp}(\mathbf{p}_k)\) denotes the index of the nonzero
entry of \(\mathbf{p}_k\). For each candidate \(n\in\mathcal{N}_l\), we set
\(\mathbf{p}_l=\mathbf{e}_n\), where \(\mathbf{e}_n\) is the \(n\)-th
canonical basis vector. Let
\[
\bar{\mathbf{p}}_q^{(n)}
=
\begin{cases}
	\mathbf{e}_n, & q=l,\\
	\mathbf{p}_q, & q\neq l.
\end{cases}
\]
Then the corresponding minimum Euclidean distance is calculated as
$d_{\min}^{(n)}	=d_{\min}(\{\bar{\mathbf{p}}_q^{(n)}\}_{q=1}^{L},\mathbf{w}).$
If \(d_{\min}^{(n)}\le C\), the candidate \(n\) is infeasible and is
discarded. Otherwise, the corresponding power splitting ratio is
$\rho_n^\star=1-C/{d_{\min}^{(n)}}.$
The second- and fourth-order channel-gain terms are given by
$
	S_2^{(n)}
	=
	\sum_{q=1}^{L}
	|\mathbf{h}^H\bar{\mathbf{p}}_q^{(n)}|^2,
$
$
	S_4^{(n)}
	=
	\sum_{q=1}^{L}
	|\mathbf{h}^H\bar{\mathbf{p}}_q^{(n)}|^4.
$
Accordingly, the objective value associated with candidate \(n\) is
evaluated as
\begin{align}
	F_n
	=
	\eta\left[
	\frac{k_2P_s}{L}\rho_n^\star S_2^{(n)}
	+
	\frac{k_4\xi P_s^2}{L}(\rho_n^\star)^2 S_4^{(n)}
	\right].
\end{align}
The optimal update of the \(l\)-th selection vector is therefore given by
\begin{align}
	\mathbf{p}_l^\star=\mathbf{e}_{n^\star},
\end{align}
where 
$	n^\star
	=
	\arg\max_{n\in\mathcal{N}_l,\ d_{\min}^{(n)}>C}F_n.$
By sequentially updating \(\mathbf{p}_1,\mathbf{p}_2,\ldots,\mathbf{p}_L\),
the proposed BCD procedure monotonically improves the reduced harvested
power \(\widetilde{E}_r\). Since the number of feasible port selection
patterns is finite and the objective value is upper bounded, the BCD
procedure converges to a coordinate-wise locally optimal solution of (P6).

\begin{algorithm}[t]
	\renewcommand{\algorithmicrequire}{\textbf{Input:}}
	\renewcommand{\algorithmicensure}{\textbf{Output:}}
	\caption{Proposed AO-Based Joint Optimization Algorithm for solving (P3)}
	\label{alg:overall_AO}
	\begin{algorithmic}[1]
		\REQUIRE \(\mathbf{h}\), \(M\), \(L\), \(P_s\), \(\sigma_z^2\), \(\gamma_{th}\), \(I_{\max}\), \(\varepsilon\), initial feasible \(\{\mathbf{p}_l^0\}\), \(\mathbf{w}^0\in\mathcal{M}\), calculate \(
		\widetilde{E}_r^{0}
		\).
		\ENSURE \(\{\mathbf{p}_l^\star\}\), \(\mathbf{w}^\star\), \(\rho^\star\).
		
		\STATE Initialize \(i=0\).
		\REPEAT
		\STATE Fix \(\{\mathbf{p}_l^{i}\}\), and solve problem (P5) via Algorithm~\ref{alg:RALM_RCG} to obtain \(\mathbf{w}^{i+1}\).
		\STATE Fix \(\mathbf{w}^{i+1}\), and solve problem (P6) via the proposed BCD method until convergence to obtain \(\{\mathbf{p}_l^{i+1}\}\).
		\STATE Update	
		\(\rho^{i+1}
		=
		 1- \frac{C}
		{d_{\min}(\{\mathbf{p}_l^{i+1}\},\mathbf{w}^{i+1})} .\)
		\STATE Compute	
		\(
		\widetilde{E}_r^{i+1}
		\)
		\STATE \(i\leftarrow i+1\).
		\UNTIL{$|\widetilde{E}_r^{i}-\widetilde{E}_r^{i-1}|\le \varepsilon$ or $i\ge I_{\max}$}
		
		\STATE \textbf{return} \(\{\mathbf{p}_l^\star\}=\{\mathbf{p}_l^{i}\}\), \(\mathbf{w}^\star=\mathbf{w}^{i}\), \(\rho^\star=\rho^{i}\).
	\end{algorithmic}
\end{algorithm}

\subsection{Overall solution}
We summarize the detailed procedures of the proposed overall solution in Algorithm \ref{alg:overall_AO}.

\subsubsection{Convergence analysis}
In Algorithm \ref{alg:overall_AO}, each subproblem is solved to a local optimum, which ensures that the objective value is nondecreasing over the iterations. Moreover, due to the finite transmit power, the objective function is upper bounded. Therefore, Algorithm \ref{alg:overall_AO} converges to a coordinate-wise locally optimal solution.

\subsubsection{Complexity Analysis}

The computational complexity of the proposed algorithm mainly arises from solving subproblems (P5) and (P6) within the AO framework. Let \(I_{\rm AO}\), \(I_{\rm out}\), \(I_{\rm in}\), and \(I_{\rm BCD}\) denote the numbers of outer AO iterations, outer RALM iterations, inner RCG iterations, and BCD iterations, respectively. For subproblem (P5), since the number of Euclidean-distance constraints is \(ML(ML-1)/2\), each inner RCG iteration requires traversing all constraint terms when evaluating the augmented Lagrangian function and its gradients, leading to a complexity of \(\mathcal{O}(M^2L^2)\). Hence, the total complexity of solving (P5) is \(\mathcal{O}(I_{\rm out}I_{\rm in}M^2L^2)\). For subproblem (P6), updating one port selection vector requires searching over at most \(N-L+1\) feasible candidate ports. The dominant cost comes from updating the minimum Euclidean distance over the affected symbol pairs, which incurs complexity \(\mathcal{O}(LM^2)\) for each candidate. Therefore, one update of \(\mathbf{p}_l\) requires complexity \(\mathcal{O}(NLM^2)\), and one BCD sweep has complexity \(\mathcal{O}(NL^2M^2)\). Thus, the total complexity of solving (P6) is \(\mathcal{O}(I_{\rm BCD}NL^2M^2)\). Combining the above two parts, the overall complexity of the proposed AO algorithm is \(\mathcal{O}\!\left(I_{\rm AO}\left[I_{\rm out}I_{\rm in}M^2L^2 + I_{\rm BCD}NL^2M^2\right]\right)\).

\section{Simulation Results}\label{section:sim}

\begin{table}
	\centering
	\caption{Simulation and Algorithm Parameters}
	\label{tab0}
	\renewcommand{\arraystretch}{1.05}
	\setlength{\tabcolsep}{10pt}
	\begin{tabular}{cc|cc}
		\toprule
		\textbf{Parameter} & \textbf{Value} & \textbf{Parameter} & \textbf{Value} \\
		\midrule
		$N_1=N_2$          & 8        & $K_{\max}$             & 30     \\
		$W_1=W_2$          & 0.5      & $R_{\max}$             & 200    \\
		$\lambda$          & 6 cm     & $I_{\max}$             & 20     \\
		$P_s$              & 30 dBm   & $\varepsilon_{\mathrm{out}}$ & $10^{-7}$ \\
		$\sigma_z^2$       & -50 dBm & $\varepsilon_{\mathrm{in}}$  & $10^{-6}$ \\
		$\eta$             & 0.9      & $\varepsilon$          & $10^{-6}$ \\
		$\epsilon_{th}'$            & 1e-3      & $\tau$                 & 0.5    \\
		$\zeta$            & 1.5      & $[\alpha_{\min},\alpha_{\max}]$                 & [0,100] 
		   \\
		$k_2$            & 0.17      & $k_4$               &  957.25  \\   
		\bottomrule
	\end{tabular}
\end{table}

In this section, simulation results are provided to verify the performance of the proposed enhanced FIM scheme as well as the effectiveness of the proposed AO algorithm. 
\subsection{Simulation Setup and Parameters}
The transmitter-to-receiver distance is set to \(10\) m. The reference attenuation at \(1\) m is assumed to be \(30\) dB, and the path-loss exponent is chosen as \(2.2\). Unless otherwise specified, the FA parameters are set to \(N=N_1\times N_2=8\times 8=64\) and \(W=W_1=W_2=0.5\). The transmit power and the noise power are fixed at \(30\) dBm and \(-50\) dBm. For the non-linear energy harvesting circuits, we set $k_2=0.17$ and $k_4=957.25$\cite{7547357}. The simulation consists of \(10^3\) Monte Carlo trials. The corresponding system and algorithm parameters are summarized in Table \ref{tab0}.

\subsection{Benchmark Schemes}

To demonstrate the effectiveness of the proposed enhanced FIM scheme, the following benchmark schemes are considered for comparison.

\begin{itemize}
	\item \textbf{Fixed FIM:} In this scheme, the ports used for FIM are fixed according to their geometric locations. Specifically, to reduce the spatial correlation among the selected ports, the \(L\) ports with the largest pairwise Euclidean distances are chosen for index modulation.
	
	\item \textbf{Top-\(L\) FIM:} In this scheme, the \(L\) ports with the strongest channel gains are selected for index modulation \cite{10440042}. Although this strategy is beneficial for energy transfer, the selected ports are usually highly correlated, which significantly degrades the information transmission performance.
	
	\item \textbf{Group FIM:} In this scheme, all available ports are uniformly divided into \(L\) groups, and the port with the strongest channel gain is selected from each group for index modulation \cite{10653737,11367092}. Compared with the Top-\(L\) FIM scheme, this grouping strategy can effectively reduce the correlation among the selected ports and thereby improve the data transmission performance. However, such a constraint on port selection may sacrifice part of the energy transfer performance.
	
	\item \textbf{FPA:}
	In this benchmark, a conventional FPA array is employed to perform transmit antenna spatial modulation (SM), where \(L\) out of \(N_t\) antennas are selected. In each transmission slot, one antenna is activated to convey additional \(\log_2(L)\) bits through the antenna index. For a fair comparison, both antenna selection and phase optimization are also considered in this scheme.
\end{itemize}

\subsection{Performance Evaluation}
\begin{figure}[t]
	\centering
	\includegraphics[width=0.9\linewidth]{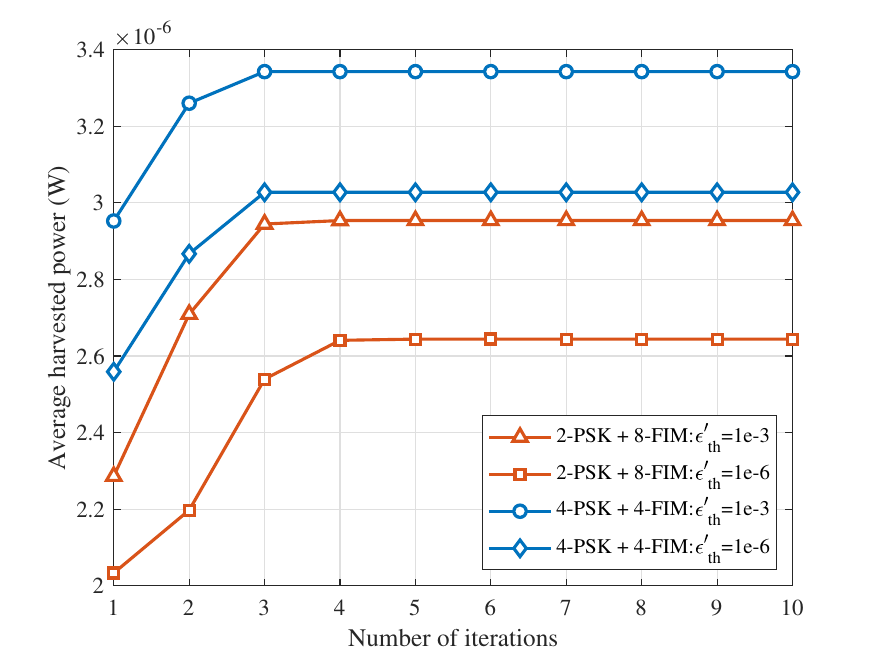}
	\caption{Convergence rate under different system setups.}
	\label{fig:WET_00}
\end{figure}
Fig. \ref{fig:WET_00} investigates the convergence of the proposed AO algorithm under different modulation schemes and BER thresholds. It can be observed that the average harvested power increases monotonically with the number of iterations and reaches a stable value within 3 to 4 iterations, which demonstrates the fast convergence behavior of the proposed algorithm. Moreover, for a given modulation scheme, a looser BER threshold leads to higher harvested power. This is because a larger BER threshold relaxes the BER constraint, thereby providing more flexibility for port selection and precoding design to enhance the WET performance. 

\begin{figure}[t]
	\centering
	\includegraphics[width=0.9\linewidth]{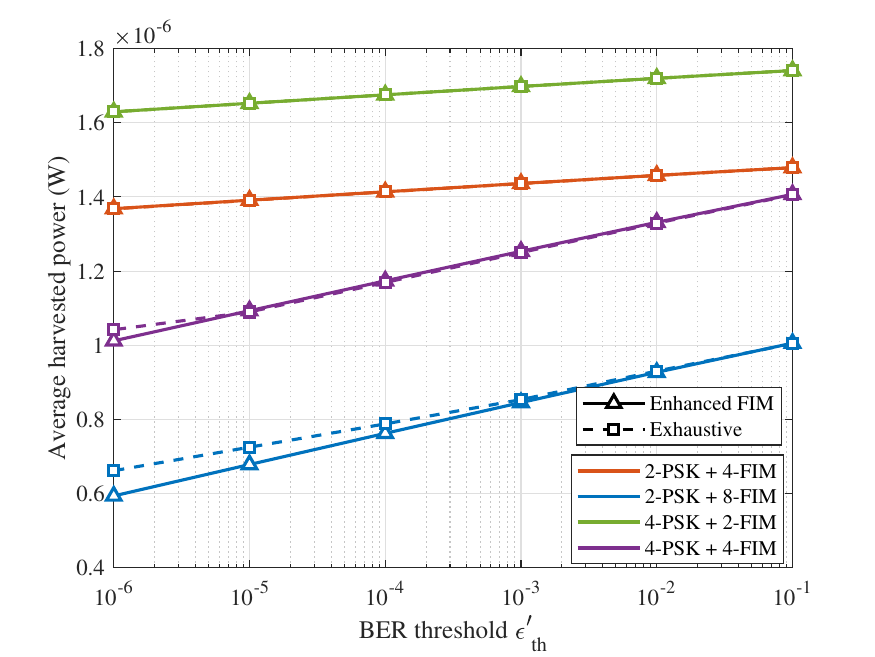}
	\caption{Average harvested power comparison of our algorithm and exhaustive search ($N_1\times N_2=4\times 4$).}
	\label{fig:WET_0}
\end{figure}
Fig. \ref{fig:WET_0} compares the average harvested power achieved by the proposed AO algorithm and the exhaustive search method under different BER thresholds and modulation schemes. It can be observed from Fig. \ref{fig:WET_0} that the proposed AO algorithm achieves performance very close to that of exhaustive search for all considered schemes, which demonstrates its near-optimality. The performance gap is particularly small for the 4-PSK+2-FIM and 2-PSK+4-FIM schemes, especially when the BER threshold is relatively high. This is because, when the modulation order is low and/or the BER threshold is loose, the BER constraint becomes less restrictive. As a result, both the proposed AO algorithm and the exhaustive search method tend to select the ports with the strongest channel magnitudes, leading to nearly identical harvested power performance. This result verifies the effectiveness of the proposed AO algorithm in achieving high harvested energy with much lower complexity than exhaustive search.

\begin{figure}[t]
	\centering
	\includegraphics[width=0.9\linewidth]{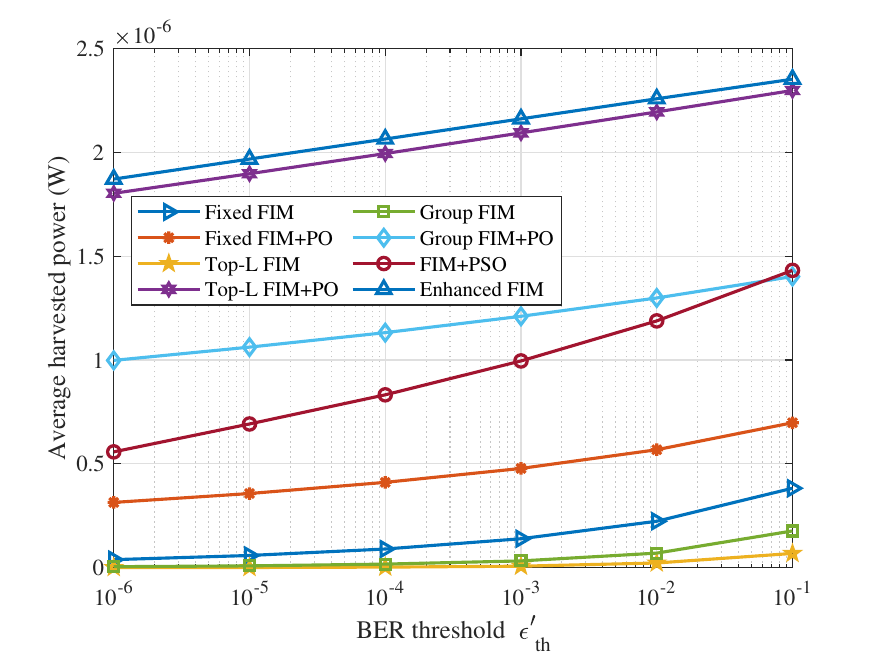}
	\caption{Average harvested power comparison of different FIM schemes.}
	\label{fig:WET_3}
\end{figure}
Fig. \ref{fig:WET_3} compares the average harvested power of different FIM schemes, where phase optimization (PO) and port selection optimization (PSO) are also taken into account. The modulation scheme is set to 2-PSK combined with 8-FIM. Observe from Fig. \ref{fig:WET_3} that the average harvested power increases as the BER threshold becomes looser. This is because a higher BER threshold relaxes the information transmission requirement and provides more freedom for energy transfer optimization. It is also observed that phase optimization can significantly improve the harvested power performance for all benchmark schemes. This is because PO adjusts the transmit phases to increase the minimum Euclidean distance of constellation, which improves the power splitter ratio $\rho$ for energy transfer. In addition, FIM with PSO consistently outperform those without PSO. This is because port selection allows the transmitter to choose more favorable ports with stronger channel gains and greater minimum Euclidean distance, thereby improving the energy transfer performance. Due to the strong correlation between ports, the Top-\(L\) FIM scheme achieves the poorest performance.
Compared with Fixed FIM, Top-\(L\) FIM, and Group FIM, the proposed enhanced FIM scheme always achieves the best harvested power performance over the whole BER range. This result shows that jointly optimizing port selection and phase shifts can explicitly control the correlation among the selected ports and better balance the trade-off between information transmission and energy transfer.

\begin{figure}[t]
	\centering
	\includegraphics[width=0.9\linewidth]{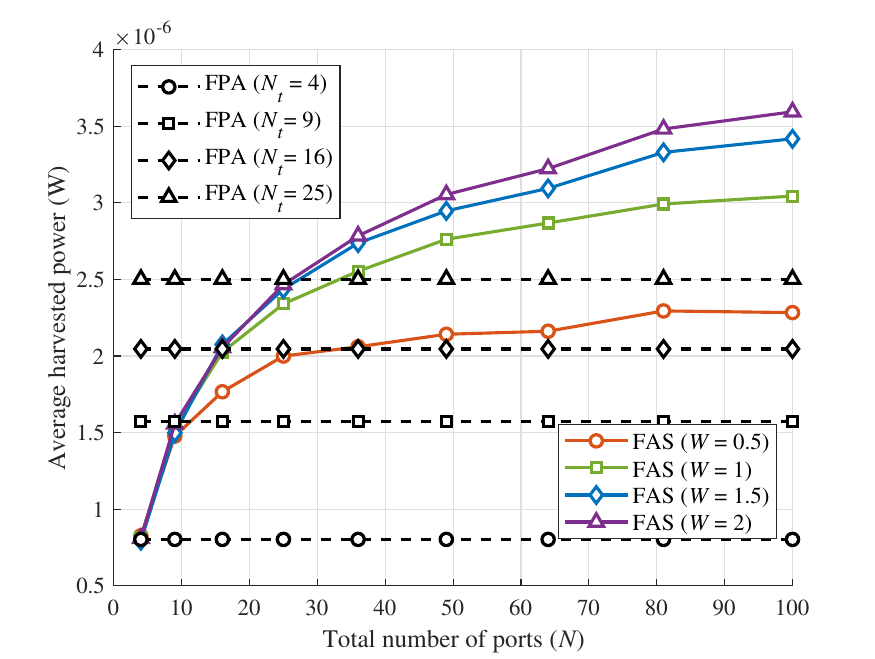}
	\caption{Average harvested power versus port number $N$ with different antenna scaling constant $W$.}
	\label{fig:WET_1}
\end{figure}

In Fig. \ref{fig:WET_1}, the average harvested power is evaluated with respect to the total number of ports \(N\) under different antenna sizes. For comparison, the average harvested power achieved by the FPA-SM scheme is also included under the same antenna size setting, where $N_t$ denotes the number of transmit antennas and the antenna spacing is $\lambda/2$. The modulation scheme is set to 4-PSK combined with 4-FIM. The BER threshold is set as 1e-3. As observed from Fig. \ref{fig:WET_1}, the average harvested power increases with the number of available ports. This is because a larger number of ports provides more spatial degrees of freedom, which enlarges the candidate set for port selection and thus increases the probability of choosing ports with stronger channel gains. In addition, a larger antenna size leads to better harvested power performance. The reason is that a larger aperture provides a wider spatial sampling region, which increases the spatial separation among candidate ports and thereby reduces the channel correlation between them. As a result, the channel responses across different ports become more distinguishable, leading to higher channel-gain diversity. Furthermore, compared with the conventional FPA-SM scheme, the proposed FIM scheme consistently achieves superior harvested power performance under the same antenna size when the number of ports in the FAS exceeds the number of antennas in the FPA. This is because the FAS provides a larger set of candidate antenna positions, offering more spatial degrees of freedom and better channel diversity for port selection.

\begin{figure}[t]
	\centering
	\includegraphics[width=0.9\linewidth]{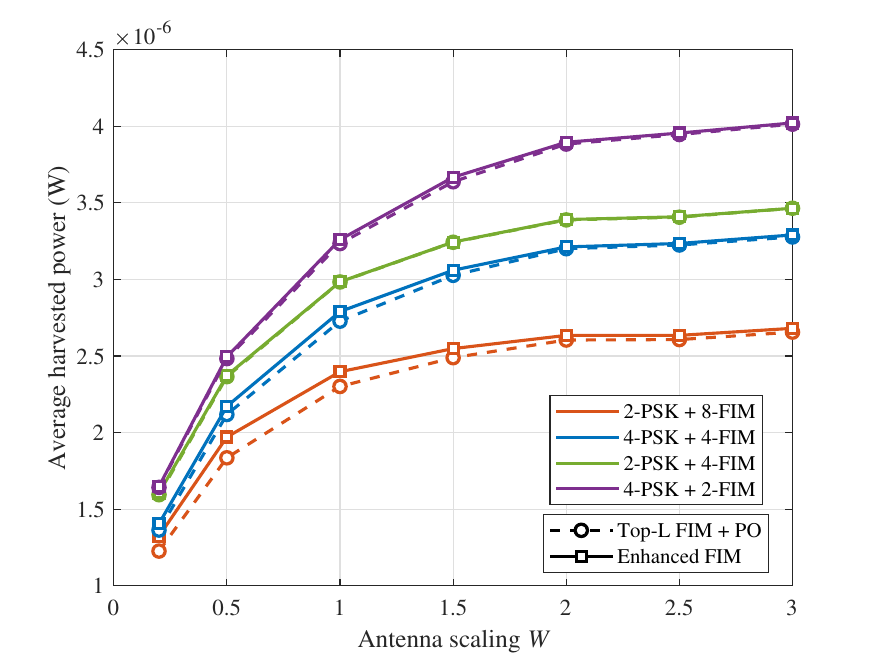}
	\caption{Average harvested power versus antenna scaling constant $W$.}
	\label{fig:WET_4}
\end{figure}
Fig. \ref{fig:WET_4} depicts the average harvested power versus the antenna scaling constant \(W\) under different modulation schemes. The proposed enhanced FIM scheme is compared with the Top-\(L\) FIM+PO scheme. For all schemes, the harvested power increases with \(W\), since a larger FA aperture enlarges the spatial sampling region, reduces inter-port correlation, and provides more opportunities to select ports with strong channel gains. The proposed enhanced FIM scheme consistently outperforms Top-\(L\) FIM+PO over the whole range of \(W\), demonstrating the benefit of jointly exploiting channel strength and inter-port correlation. When the modulation order is low, the gap becomes small because the BER constraint is easier to satisfy, and the proposed design tends to approach energy-oriented port selection.

\begin{figure}[t]
	\centering
	\includegraphics[width=0.9\linewidth]{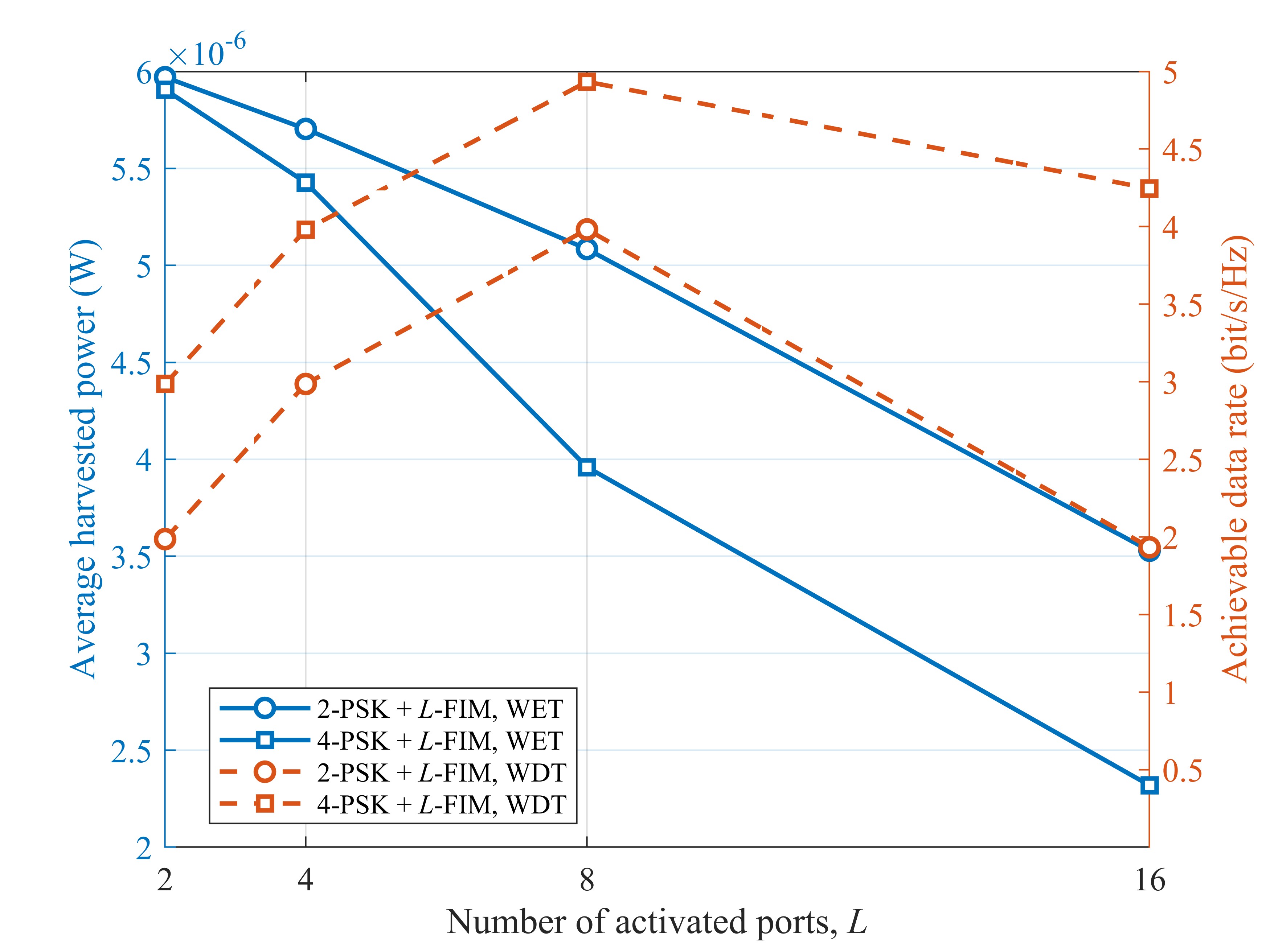}
	\caption{Average harvested power and achievable data rate versus number of activated ports $L$ .}
	\label{fig:WET_WDT_L}
\end{figure}
Fig. \ref{fig:WET_WDT_L} illustrates the average harvested power and achievable data rate versus the number of activated ports $L$ under different modulation schemes. It is observed that the harvested power decreases with increasing $L$, since activating more ports introduces more information in FIM index domain and tightens the BER constraint, thereby reducing the flexibility of port selection for energy harvesting. In contrast, the achievable data rate first increases and then decreases as $L$ grows. This is because a larger $L$ provides more index bits, while an excessively large $L$ reduces the minimum Euclidean distance among transmit candidates and degrades the BER performance. Hence, there exists a proper choice of $L$ that balances WET and WDT performance.

\begin{figure}[t]
	\centering
	\includegraphics[width=0.9\linewidth]{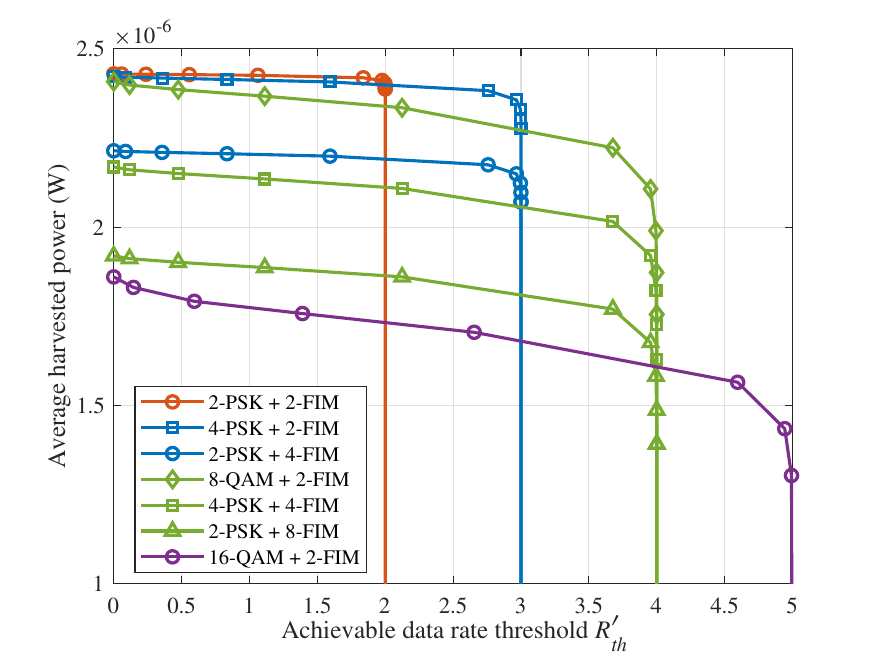}
	\caption{Average harvested power versus different achievable data rate threshold $R_{th}'$ .}
	\label{fig:WET_2}
\end{figure}
Fig. \ref{fig:WET_2} illustrates the R-E trade-off under various modulation schemes. 
It can be observed from Fig. \ref{fig:WET_2} that the average harvested power consistently decreases as the data rate threshold increases across all schemes. This is because a higher data rate imposes a tighter BER constraint, which shrinks the feasible design space for port selection and precoding. Consequently, the system has fewer degrees of freedom to maximize energy transfer. In addition, we evaluate schemes operating at the same total transmission rate. The results show that allocating more bits to conventional constellation modulation, rather than the FIM domain, yields better harvested power. For instance, at a total rate of 3 bits/symbol, the 4-PSK+2-FIM scheme outperforms 2-PSK+4-FIM. This is because increasing the FIM order requires activating more candidate ports for index modulation, which weakens the flexibility of selecting the strongest ports for energy transfer and thus leads to a larger performance loss in harvested power.

\begin{figure}[t]
	\centering
	\includegraphics[width=0.9\linewidth]{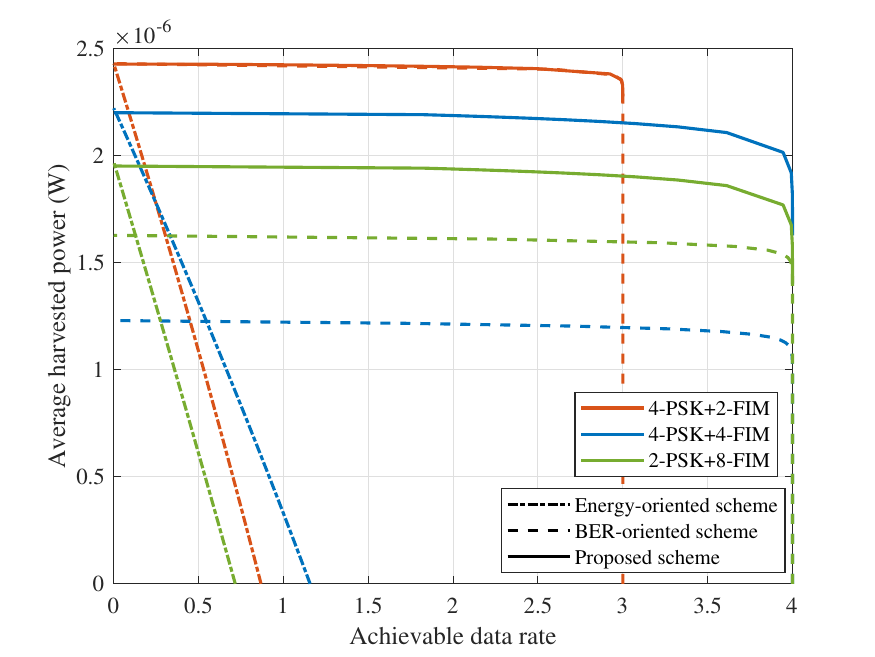}
	\caption{Average harvested power versus different achievable data rate.}
	\label{fig:WET_5}
\end{figure}
Fig. \ref{fig:WET_5} compares the R-E regions achieved by the energy-oriented scheme, the BER-oriented scheme, and the proposed enhanced FIM scheme. Here, the energy-oriented scheme maximizes the average harvested energy without considering the BER constraint, whereas the BER-oriented scheme maximizes the minimum Euclidean distance without taking the harvested energy into account. Specifically, the energy-oriented scheme is realized by selecting the \(L\) ports with the strongest channel amplitudes. For the BER-oriented scheme, the port selection vectors and precoder are optimized by solving a minimum-Euclidean-distance maximization problem using the proposed AO framework. The corresponding R--E regions of these schemes are then obtained by varying the power splitting ratio \(\rho\) from \(0.001\) to \(0.999\).
As shown in Fig. \ref{fig:WET_5}, the proposed enhanced FIM scheme achieves a larger R-E region by jointly considering harvested energy and BER performance. The energy-oriented scheme attains high harvested power but suffers from poor achievable rate due to highly correlated selected ports, while the BER-oriented scheme improves communication performance at the cost of harvested power. For 4-PSK+2-FIM, the proposed and BER-oriented curves nearly overlap because the BER constraint is easier to satisfy and both schemes tend to select ports with strong channel gains.
\section{Conclusion}\label{section:conclusion}
In this paper, we studied an enhanced FIM-assisted IDET system, where the information bits are jointly transmitted through the conventional amplitude-phase domain and the fluid index domain. Closed-form expressions for the average harvested power and tractable BER/rate bounds were derived. In order to enhance the WDT and WET performance, a joint optimization problem was formulated to maximize the average harvested power under BER and achievable rate constraints by jointly optimizing the port selection, precoding vector, and power splitting ratio. An AO algorithm was then developed, where the precoding and port selection subproblems were solved by the RALM and BCD algorithms, respectively. Simulation results showed that the proposed scheme achieves better WET and WDT performance than the benchmark schemes, while the proposed AO algorithm provides near-optimal performance with much lower complexity than exhaustive search. In addition, the results reveal that a larger antenna region and a greater number of ports can achieve better performance. Therefore, the proposed enhanced FIM-assisted IDET design offers an effective way to exploit the spatial flexibility of fluid antenna systems for simultaneous information and energy transfer.
\bibliography{Reference}

\end{document}